\documentstyle{article}
\font\eurm=eurm10       
\newcommand{\Eu}[1]{\mbox{\eurm #1}}
\font\msam=msam10       

\font\msbm=msbm10 scaled 1000
\font\msbmsmall=msbm10 scaled 800
\font\msbmtiny=msbm10 scaled 700
\newfam\msbmfam
\textfont\msbmfam=\msbm
\scriptfont\msbmfam=\msbmsmall
\scriptscriptfont\msbmfam=\msbmtiny
\def\Bbb{\fam\msbmfam\msbm}
\def\proof{\par{\it Proof.\/} \ignorespaces}
\def\endproof{{\msam\char'3}}

\newtheorem{theorem}{Theorem}[section]

\newtheorem{corollary}[theorem]{Corollary}

\def\Re{\mbox{\rm Re}\, }
\def\Im{\mbox{\rm Im}\, }
\def\diag{\mbox{\rm diag}}

\def\varkappa{\mbox{\msbm \char'173}}      
\def\scriptvarkappa{\mbox{\msbmsmall \char'173}}      
\begin{document}
\centerline{\Large Monodromy deformation approach to the scaling}
\centerline{\Large limit of the Painlev\'e first equation}

\bigskip
\centerline{\large Andrei A.~Kapaev}
\centerline{\it St.~Petersburg Department}
\centerline{\it of Steklov Mathematical Institute,}
\centerline{\it Fontanka 27, St.~Petersburg, 191011, Russia}
\centerline{\it E-mail: kapaev@pdmi.ras.ru}

\bigskip
\begin{abstract}
The isomonodromy deformation equation for a $2\times2$ matrix linear ODE with 
a large parameter can be locally reduced to a (hyper)elliptic equation. To 
globalize this result, we apply the isomonodromy deformation method and 
obtain the modulation equations for the asymptotic algebraic curve. The 
method is applied to the degenerate solution of the Painlev\'e first equation.
\end{abstract}

\section{Introduction}

The classical Painlev\'e equations are found by P.~Painlev\'e and B.~Gambier 
\cite{painleve1, gambier} as the only up to M\"obius transformations 
irreducible second-order ODEs $y_{xx}=R(x,y,y_x)$ with the rational in $y_x$, 
algebraic in $y$ and analytic in $x$ r.h.s.\ whose general solutions are free 
from movable critical points. The last condition is usually referred to as 
the Painlev\'e property while the equations which enjoy this property are 
called the Painlev\'e-type equations. The term {\it irreducible} means that 
the generic solutions of the Painlev\'e equations can not be expressed in 
terms of elementary or classical special functions and hence deserve the name 
of {\it Painlev\'e transcendents}. 

The main tool in the classical works \cite{painleve1, gambier} is the 
$\alpha$-test of Painlev\'e based on the assertion that the limiting form of 
the scaled Painlev\'e-type equation, or the {\it reduced form}, also 
possesses the Painlev\'e property \cite{ince}. E.g., the Painlev\'e second 
equation PII, 
\begin{equation}\label{PII}
y_{xx}=2y^3+xy-\alpha,
\end{equation}
scaled in accord with the relations
\begin{equation}\label{2scl6}
x=\delta^{-2/3}x_0+\delta^{1/3}t,\quad
y=\delta^{-1/3}v,\quad
y_x=\delta^{-2/3}v_t,\quad
\alpha=\delta^{-1}\beta_0+\beta_1,
\end{equation}
where $t$ is the new independent variable, $v$ is the new dependable, and
$x_0$, $\beta_0$, $\beta_1$ are some constant parameters, becomes
\begin{equation}\label{2scl7}
v_{tt}=2v^3+x_0v-\beta_0+\delta(tv-\beta_1).
\end{equation}
At $\delta=0$, the latter is known to have the Painlev\'e property since it 
can be integrated by the use of elliptic functions. If $x_0=-6$ and 
$\beta_0=-4$, the additional scaling changes $t=\delta^{-1/5}\tau$, 
$v=1+\delta^{2/5}z$, $\beta=-\delta^{-1/5}\gamma$, turn equation 
(\ref{2scl7}) into $z_{\tau\tau}=6z^2+\tau+\gamma+\delta^{2/5}(2z^3+\tau z)$, 
which, at $\delta=0$, coincides with the Painlev\'e first equation PI.

The scaling reduction of the Painlev\'e equation to the elliptic equation 
means that the Painlev\'e function is described {\it locally} by the
elliptic function. This fact was justified for equations PI and PII by 
P.~Boutroux \cite{boutroux} and later by P.~Doran-Wu and N.~Joshi 
\cite{doran-wu_joshi, joshi}. In our terms, P.~Doran-Wu and N.~Joshi studied 
the local behavior of the function $v(t)$ ($x_0=0$) with the initial 
condition $(t,v,v_t)=(t_0,\eta,\eta')$ where $t_0,\eta,\eta'$ are such 
constants that
$D:=\eta'^2-\eta^4+2\beta_0\eta\neq3(\beta_0/2)^{4/3}e^{i\frac{2\pi}{3}k}$,
$k\in{\Bbb Z}$. They proved that, in the region $|t-t_0|<|\ln\delta|$, in the 
leading order w.r.t.\ $\delta\to0$, the function $v(t)$ is uniformly 
approximated by the elliptic function $V(t)$, $V'^2=V^4-2\beta_0V+D$. 

Any extension of the above result beyond the indicated boundary involves 
{\it modulation} of the elliptic curve, i.e.\ certain change of $D$ when 
$x_0$ in (\ref{2scl6}) varies. For the first time, the modulation equations 
for the elliptic asymptotics as $|x|\to\infty$ of the Painlev\'e first and 
second functions were found by P.~Boutroux \cite{boutroux}. Namely, letting 
in (\ref{PII}) $y=\sqrt{x}v$, $t=\frac{2}{3}x^{3/2}$ and integrating, he has 
found the equation $(v')^2=v^4+v^2+D(t)$, where $D(t)$ is a transcendent 
function. Using integral estimates, P.~Boutroux obtained the variation of 
$D(t)$ on a period $\omega_k=\oint_{\cal L}(V^4+V^2+D_k)^{-1/2}\,dV$ of the 
local elliptic ansatz, $(V')^2=V^4+V^2+D_k$, $D_k=const$, and has shown that, 
along the unbounded ``line of periods", there exists 
$\lim_{k\to\infty}D_k=D_{\infty}$. The latter satisfies the transcendent 
{\it Boutroux equation} 
\begin{equation}\label{Boutroux_eq}
\oint_{\cal L}\sqrt{V^4+V^2+D_{\infty}}\,dV=0.
\end{equation}
Nowadays it is clear that (\ref{Boutroux_eq}) is strongly related to the 
Hamiltonian structure associated to the Painlev\'e equation and appears as 
the asymptotic form of the {\it relative Poincar\'e invariant} \cite{arnold} 
for PII.

Another approach to the modulation equations is provided by the method of 
G.~Kuzmak \cite{kuzmak} who proposed to expand $v$ in an ascending series in 
$\delta$, $v=V+\delta v_1+\dots$\ with a periodic leading term $V$, and to 
introduce, aside from the ``quick" variable $t$, another, formally 
independent, ``slow" variable $T=\delta t$ for description of the variation
of the period and phase shift in $V(t,T)$. Eliminating ``secular" terms from 
the linear differential equation for $v_1$, one finds an integral condition 
for $V(t,T)$ which is usually interpreted as the constancy of the Lagrange's
action. This condition is the desired modulation equation for $D(T)$. Using 
this way, N.~Joshi and M.~Kruskal \cite{joshi_kruskal} described the 
modulation of the elliptic asymptotic solutions of PI and PII w.r.t.\ 
$\varphi=\arg x$ as $|x|\to\infty$. 

Applying the Kuzmak's method and using the connection formulae of 
M.~Ablow\-itz and H.~Segur \cite{abl_seg2, seg_abl} for the class of the 
physically interesting decreasing as $x\to+\infty$ solutions of PII 
($\alpha=0$), $y(x,a)\sim a\,{A\!i}(x)\sim
\frac{a}{2\sqrt\pi}x^{-1/4}e^{-\frac{2}{3}x^{3/2}}$,
J.~Miles \cite{miles} described the asymptotic dependence of the solution on
the amplitude $a$ as $a\to i\infty$ and $a\to1-0$.

Using the same ideas, O.~Kiselev \cite{kiselev} constructed the imaginary on 
the real axis solution of PII ($\alpha\to i\infty$) which is monotonic as 
$x\to+\infty$ and oscillates as $x\to-\infty$. The oscillating tail is 
described by the modulated elliptic function, $V'^2=V^4+x_0V^2+2iV+D$, where 
$D$ depends on $x_0$ in such a way that 
$\oint\sqrt{V^4+x_0V^2+2iV+D}\,dV\equiv2\pi$.

An equivalent way leading to the modulation equations in a differential form 
is described by J.~Whitham \cite{whitham}. This approach utilizes averaging 
w.r.t.\ the ``fast" variable rather than considers a single period of the 
elliptic function. This idea was applied by V.~Novokshenov \cite{nov1, nov2}
and V.~Vereshchagin \cite{vereschagin1, vereschagin2} to the asymptotic as 
$|x|\to\infty$ description of equations PII, PIII and PIV, and to description of 
the scaling limits in the classical Painlev\'e equations. The differential
modulation equations for the scaling limits are found in the explicit form 
$\partial_TD=f(D,K,E)$, where $T$ is the ``slow" variable, $D$ is the
module, and $K$, $E$ are the complete elliptic integrals.

In the most systematic way, the averaging method for the integrable 
``soliton" equations was developed by H.~Flaschka, M.~Forest and 
D.~McLaughlin \cite{FFM} and I.~Krichever \cite{krichever}. Applications of 
their ideas to equation PI $y_{xx}=6y^2+x$ are presented in the papers of 
I.~Krichever \cite{krichever2}, S.~Novikov \cite{novikov} and F.~Fucito, 
A.~Gamba, M.~Martellini and O.~Ragnisco \cite{FGMR}. In particular, the 
differential Whitham equations for PI imply the integral conditions for the 
asymptotic elliptic curve \cite{krichever2},
\begin{equation}\label{normaliz}
\Im\oint_{a,b}w(\lambda)\,d\lambda=h_{a,b}=const,\quad
w^2=4\lambda^3+2x\lambda-g_3,
\end{equation}
which yield the invariant $g_3$ in the ansatz $y(x)=\wp(x-x_0;-2x,g_3(x))$ 
as a transcendent function of $x$.

Possibly, the most effective approach to the Painlev\'e equations is based on 
their isomonodromy interpretation. R.~Fuchs \cite{fuchs} constructed a linear 
scalar second-order ODE with four Fuchsian singularities and one apparent 
singular point whose isomonodromy deformation is governed by the Painlev\'e 
sixth equation PVI. R.~Garnier \cite{garnier} has found the linear equations 
associated to the lower Painlev\'e equations imposing on the equation of 
R.~Fuchs the scaling changes of variables of P.~Painlev\'e \cite{painleve2} 
for the cascade of the successive reductions of the Painlev\'e equations to 
each other. These linear equations also form a cascade of reductions arising 
in some neighborhoods of the merging singularities and turning points. 
Comments and some technical details missed in the original paper of 
R.~Garnier can be found in the article of P.~Boutroux \cite{boutroux}. 
L.~Schlesinger \cite{schlesinger} developed the isomonodromy deformation 
theory for generic linear matrix equations with Fuchsian singular points, 
while R.~Garnier \cite{garnier3} transformed his scalar equations to the 
matrix form and discovered the link between the isomonodromy and 
``isospectral" deformations \cite{garnier2}. In the papers of M.~Jimbo,
T.~Miwa and K.~Ueno \cite{jmu} and H.~Flaschka and A.~Newell \cite{fn} the
isomonodromy deformation theory was developed for the equations with
non-Fuchsian singularities.

The asymptotics of the Painlev\'e functions as $x$ tends to one of its fixed 
singularities were studied by M.~Jimbo, A.~Its, V.~Novokshenov, B.~McCoy, 
Sh.~Tang, B.~Suleymanov, A.~Kitaev, A.~Kapaev and others, see \cite{jimbo, 
its_nov, mccoy_tang, suleiman, kapaevP2sng, kitaev0}. The modulation 
equations for the elliptic asymptotic solutions as $|x|\to\infty$ were 
discussed from the isomonodromy deformation viewpoint in \cite{nov1, 
kapaevP2ell, kapaevP2P4, kap_kit2, nov2}. The majority of the mentioned 
authors used the WKB method as the basic ingredient of the asymptotic 
investigation of the linear equation.

H.~Flaschka and A.~Newell \cite{fn2} observed that the WKB approximation to a 
solution of the $\lambda$-equation in the Lax pair, 
$\Psi_{\lambda}=\delta^{-1}A\Psi$, $d\Psi=U\Psi$, satisfies as $\delta\to0$ 
the ``isospectral" problem, $A\psi=\mu\psi$, $d\psi=U\psi$. Making use of 
this fact, K.~Takasaki \cite{takasaki} avoided the averaging procedure and 
proposed the differential modulation equations simply comparing the WKB 
approximation for the first equation in the Lax pair and the Baker-Akhiezer 
function for the second equation. 

The scaling limits in Painlev\'e equations find important applications in 
topological field theories and quantum gravity \cite{DS, moor, BK, GM, 
paniak_szabo}, in the theory of semi-classical orthogonal polynomials and 
random matrices \cite{freud, nevai, mehta, pastur, magnus}. In the series of 
papers \cite{fokas_its_kitaev}, A.~Fokas, A.~Its and A.~Kitaev described the 
continuous limit in the matrix model of quantum gravity as the isomonodromy 
scaling limit induced by the sequence of the B\"acklund transformations in 
PIV. Similarly, A.~Kapaev and A.~Kitaev \cite{kapaev_kitaev} described the 
isomonodromy scaling limit from PII to PI, P.~Bleher and A.~Its 
\cite{bleher_its} computed the asymptotics of the semi-classical orthogonal 
polynomials with the quartic weight related to the isomonodromy scaling limit 
in PIV, while J.~Baik, P.~Deift and K.~Johansson \cite{baik_deift_johansson} 
found the distribution law of the increasing subsequence of random 
permutations related to the isomonodromy scaling limit from PIII to PII. 
The elliptic asymptotic solutions for the isomonodromy scaling limits in PII 
and the corresponding modulation equations are found in \cite{kapaevP2scl}. All 
the modulation equations which previously appeared in the isomonodromy 
deformation context, see \cite{nov1, kapaevP2ell, kapaevP2P4, kap_kit2, nov2, 
kapaevP2scl}, have the integral form (\ref{normaliz}) with the proper 
integrands $w(\lambda)$ and the trivial r.h.s., $h_a=h_b=0$.

In Section~2, we recall the basic notions of the isomonodromy deformation 
theory of the linear systems with rational coefficients, introduce the 
scaling changes of variables which lead to the ``isospectral" problem, and 
define the WKB approximation. In Section~3, we introduce the spectral curve 
and find the modulation equations for the (hyper)elliptic asymptotic 
solutions of the isomonodromy deformation equation, describe the domain of 
their validity and discuss their solvability. In Section~4, we apply the 
method to equation PI.

\section{Isomonodromy deformations and scaling limits}

Here, we recall the basic facts of the theory of the matrix equations with 
rational coefficients following the text books of W.~Wasow \cite{wasow}, 
F.~Olver \cite{olver} and M.~Fedorjuk \cite{fedor}, and of the isomonodromy 
deformation theory following M.~Jimbo, T.~Miwa and K.~Ueno \cite{jmu, jm2} 
and H.~Flaschka and A.~Newell \cite{fn}. 

Consider a first order $2\times2$ matrix equation with rational coefficients,
\begin{equation}\label{Psi_lambda}
\frac{d\Psi}{d\lambda}=A(\lambda)\Psi,\quad
A(\lambda)=\sum_{\nu=1}^n\sum_{k=0}^{r_{\nu}}
\frac{A_{\nu,-k}}{(\lambda-a^{(\nu)})^{k+1}}-
\sum_{k=1}^{r_{\infty}}A_{\infty,-k}\lambda^{k-1},\quad
\mbox{\rm Tr}\,A(\lambda)=0.
\end{equation}
Without loss of generality, $\lambda=\infty$ is the singular point of the 
highest Poincar\'e rank, i.e.\ $r_{\infty}\geq r_{\nu}$, $\nu=1,\dots,n$.
We call equation (\ref{Psi_lambda}) {\it generic} if the eigenvalues of 
$A_{\nu,-r_{\nu}}$ are distinct for $r_{\nu}\neq0$ and if they are distinct 
modulo integers for $r_{\nu}=0$. Let the generic equation (\ref{Psi_lambda}) 
be diagonal at infinity, i.e.\ 
$A_{\infty,-r_{\infty}}=x_{-r_{\infty}}^{(\infty)}\sigma_3$ for
$r_{\infty}\geq1$, or 
$A_{\infty,0}=-\sum_{\nu=1}^nA_{\nu,0}=x_0^{(\infty)}\sigma_3$ for 
$r_{\infty}=0$, where $\sigma_3=\diag(1,-1)$. 

The generic equation (\ref{Psi_lambda}) has the formal solution near the 
singularity $a^{(\nu)}$, 
\begin{eqnarray}\label{Psi_formal}
&&\Psi_{\nu}(\lambda)=W^{(\nu)}\hat\Psi^{(\nu)}(\lambda)
e^{\theta^{(\nu)}(\lambda)\sigma_3},\quad
\nu=1,\dots,n,\infty,\nonumber
\\
&&\hat\Psi^{(\nu)}(\lambda)=I+\sum_{j=1}^{\infty}\psi_j^{(\nu)}\xi^j,\quad
\theta^{(\nu)}(\lambda)=
\sum_{j=1}^{r_{\nu}}x_{-j}^{(\nu)}\frac{\xi^{-j}}{(-j)}+
x_0^{(\nu)}\ln\xi,
\end{eqnarray}
where $\xi=\lambda-a^{(\nu)}$ for a finite singular point $a^{(\nu)}$ and 
$\xi=1/\lambda$ otherwise. The coefficients $\psi_j^{(\nu)}$,
$x_{-j}^{(\nu)}$ of the formal expansion (\ref{Psi_formal}) are determined 
uniquely by the eigenvector matrix $W^{(\nu)}$ of $A_{\nu,-r_{\nu}}$. The 
formal solution $\Psi_{\infty}(\lambda)$ is fixed by the normalization 
condition $W^{(\infty)}=I$. The eigenvector matrix $W^{(\nu)}$ for the finite 
singularity $a^{(\nu)}$ is unique modulo permutation and scaling of the 
eigenvectors. Thus, $\Psi_{\nu}(\lambda)$ (\ref{Psi_formal}) is unique modulo 
the right constant diagonal and permutation matrix multipliers. In practice, 
this ambiguity does not make any difficulty, and we assume below that the 
eigenvector matrices $W^{(\nu)}$ are chosen in some convenient way. 

If $r_{\nu}=0$, the singularity is called {\it Fuchsian}. For Fuchsian
singular point, the series $\hat\Psi^{(\nu)}(\lambda)$ converges in a disk 
centered at $a^{(\nu)}$ \cite{wasow}, and the rational terms in the sum for 
the diagonal matrix $\theta^{(\nu)}(\lambda)$ are absent. If $r_{\nu}\neq0$, 
expansion (\ref{Psi_formal}) gives an asymptotic representation for 
$\Psi(\lambda)$ in one of the {\it Stokes sectors} near $\xi=0$, i.e.\ 
for $\arg\xi\in\bigl(\frac{\pi}{r_{\nu}}(k-1)+\varphi+\varepsilon,
\frac{\pi}{r_{\nu}}(k+1)+\varphi-\varepsilon\bigr)$, where 
$\varphi=\frac{1}{r_{\nu}}(\arg x_{-r_{\nu}}^{(\nu)}-\frac{\pi}{2})$,
$k=1,\dots,2r_{\nu}$, $\varepsilon>0$. 

A comprehensive description of the fundamental solutions of the non-generic 
equation can be found in \cite{wasow}.

The ratio $\tilde\Psi^{-1}(\lambda)\Psi(\lambda)$ of any two solutions
$\Psi$ and $\tilde\Psi$ of (\ref{Psi_lambda}) does not depend on $\lambda$. 
Ratios of the fundamental solutions normalized by (\ref{Psi_formal}) are 
usually called the {\it monodromy data}. The latter include the monodromy 
matrices $M_{\nu}$ which describe the branching of $\Psi(\lambda)$ near the 
singularities, the Stokes matrices $S_{\nu}^{(k)}$, $k=1,\dots,2r_{\nu}$, 
which describe the Stokes phenomenon near the non-Fuchsian singularities, and 
the connection matrices $E_{\rho}^{\nu}$ which are the ratios of the 
fundamental solutions normalized at distinct singular points. At this stage, 
we do not need more detailed description of the monodromy data, see 
\cite{jmu, fn}. 

The set of {\it deformation parameters} is specified for generic equation 
(\ref{Psi_lambda}) in \cite{jmu}. These include the positions $a^{(\nu)}$ of the 
singular points and the coefficients $x_{-j}^{(\nu)}$, $j\neq0$, for
$\theta^{(\nu)}(\lambda)$ in (\ref{Psi_formal}), and form together the vector 
$x$ (the extension of this definition to some non-generic equations is given 
in \cite{fedor2}). Remaining parameters $x_0^{(\nu)}$ are usually called the 
{\it formal monodromy exponents}. 

Using the linear transformation of the complex $\lambda$-plane, it is always 
possible to fix two of the deformation parameters, e.g.\ to move two of the
finite singular points $a^{(\nu)}$ to 0 and 1. Alternatively, if
$r_{\infty}\geq2$, one may put $x_{-r_{\infty}}^{(\infty)}=1$, 
$x_{1-r_{\infty}}^{(\infty)}=0$. Thus the number of the actual deformation 
parameters is equal to 
\begin{equation}\label{number_of_parameters}
m_c=n+r-2,\quad
r=\sum_{\nu=1,\dots,n,\infty}r_{\nu}.
\end{equation}

Let $d$ denote the exterior differentiation w.r.t.\ entries of the vector of
deformation parameters $x$. The monodromy data of (\ref{Psi_lambda}) do not 
depend on $x$ if and only if there exist 1-forms $\Omega$ and 
$\Theta^{(\nu)}$ such that the fundamental solutions above additionally 
satisfy equations \cite{jmu}
\begin{equation}\label{Psi_x}
d\Psi=\Omega\Psi,\quad
dW^{(\nu)}=\Theta^{(\nu)}W^{(\nu)},\quad
\nu=1,\dots,n,\infty.
\end{equation}
The compatibility condition of (\ref{Psi_lambda}), (\ref{Psi_x}),
\begin{equation}\label{Frobenius}
dA=\frac{\partial\Omega}{\partial\lambda}+[\Omega,A],\quad
d\Omega=\Omega\wedge\Omega,
\end{equation}
is the completely integrable differential system \cite{jmu} whose fixed
singularities are the planes $a^{(\nu)}=a^{(\rho)}$, $\nu\neq\rho$, and
$x_{-r_{\nu}}^{(\nu)}=0$, $r_{\nu}\neq0$ \cite{jmu}. The generic system
(\ref{Frobenius}) corresponding to $m_c=1$ is equivalent to one of the 
classical Painlev\'e equations \cite{jm2, fn}. 

An extension of the isomonodromy deformation theory to the non-generic 
equations is given in \cite{fedor2}. We also mention an exhaustive study 
\cite{DM} of a quite important particular non-generic equation ÿÿÿÿÿÿÿ
(\ref{Psi_lambda}) with the matrix $A(\lambda)$ given by
\begin{equation}\label{A_P6}
A(\lambda)=\sum_{\nu=1}^3\frac{A_{\nu}}{\lambda-a^{(\nu)}},
\end{equation}
where the residue matrices $A_{\nu}$, $\nu=1,2,3$, are all nilpotent and 
satisfy the normalization condition $A_1+A_2+A_3=-\alpha\sigma_3$. If 
$a^{(1)}=0$, $a^{(2)}=1$, $a^{(3)}=x$, then 
$\Omega=-\frac{A_3}{\lambda-x}dx$, and the Schlesinger system which governs 
the isomonodromy deformations of (\ref{Psi_lambda}), (\ref{A_P6}) is reduced 
to the particular case of the PVI equation,
\begin{eqnarray}\label{P6}
&&y_{xx}=\frac{1}{2}\bigl(\frac{1}{y}+\frac{1}{y-1}+\frac{1}{y-x}\bigr)y_x^2-
\bigl(\frac{1}{x}+\frac{1}{x-1}+\frac{1}{y-x}\bigr)y_x+\nonumber
\\
&&+\frac{y(y-1)(y-x)}{2x^2(x-1)^2}
\Bigl((2\alpha-1)^2+\frac{x(x-1)}{(y-x)^2}\Bigr),
\end{eqnarray}
where $y$ is the only zero of the entry $(A(\lambda))_{12}$. Physical
applications of (\ref{P6}) include the self-dual Bianchi~IX model 
\cite{tod, cha, oku} and 2D-topological field theories \cite{D1}.

Let us make in (\ref{Psi_lambda}) the changes of variables, 
\begin{eqnarray}\label{scaling_change}
&&\lambda=\delta^{\scriptvarkappa}\zeta,\quad
a^{(\nu)}=\delta^{\scriptvarkappa}(b^{(\nu)}+\delta c^{(\nu)}),\quad
\varkappa=const,\quad
\nu=1,\dots,n,
\\
&&A_{\nu,-k}=\delta^{k\scriptvarkappa-1}B_{\nu,-k},\quad
A_{\infty,-k}=\delta^{-k\scriptvarkappa-1}B_{\infty,-k},\quad
k=0,\dots,r_{\nu},\nonumber
\end{eqnarray}
so that, for generic equation (\ref{Psi_lambda}),
\begin{equation}\label{x_change}
x_{-k}^{(\nu)}=\delta^{k\scriptvarkappa-1}
(t_{-k}^{(\nu)}+\delta\tau_{-k}^{(\nu)}),\quad
x_{-k}^{(\infty)}=
\delta^{-k\scriptvarkappa-1}
(t_{-k}^{(\infty)}+\delta\tau_{-k}^{(\infty)}).
\end{equation}
The constants $b^{(\nu)}$, $t_{-k}^{(\nu)}$, $k\neq0$, form the vector $t$
which mark the center of the asymptotic domain as $\delta\to+0$. Inequalities 
$|c^{(\nu)}|,|\tau_{-k}^{(\nu)}|<const$, or, in the vector form, 
$\|\tau\|<const$, give the range of the ``local" deformation. The 
entries of $t$ are called the ``slow" variables, while the entries of 
$\tau$ are usually called the ``quick" or ``fast" variables. The parameters
$t_0^{(\nu)}$ and $\tau_0^{(\nu)}$ form the vectors $\alpha$ and $\beta$.

\smallskip
{\it Remark 1}. If $r_{\infty}\neq0$ and $x_{-r_{\infty}}^{(\infty)}$ is fixed
by the use of the proper change of the independent variable $\lambda$ then 
equation (\ref{x_change}) yields $\varkappa=-1/r_{\infty}$. If all the
singularities are Fuchsian, i.e.\ $r_{\nu}=0$, $\nu=1,\dots,n,\infty$, one 
may put for simplicity $\varkappa=0$ which yields the famous autonomous 
Garnier system \cite{garnier2}. E.g., using in (\ref{P6}) the changes 
$x=t+\delta\tau$, $\alpha=\delta^{-1}\alpha_0+\alpha_1$, $\delta\to0$, we 
obtain the reduced autonomous equation 
$y_{\tau\tau}=
\frac{1}{2}\bigl(\frac{1}{y}+\frac{1}{y-1}+\frac{1}{y-t}\bigr)y_{\tau}^2+
2\alpha_0^2\frac{y(y-1)(y-t)}{t^2(t-1)^2}$, which has the first 
integral
\begin{equation}\label{P6_integral}
D_0=\frac{t^2(t-1)^2}{4y(y-1)(y-t)}y_{\tau}^2-\alpha_0^2y,
\end{equation}
and therefore is solvable in terms of elliptic functions.

\smallskip
{\it Remark 2}. Interpreting the change of the formal monodromy exponents 
$x_0^{(\nu)}$ as the result of the Schlesinger transformation action which
preserve all the monodromy data \cite{jm2, fokas_its_kitaev}, we arrive at 
the quantization condition,
$x_0^{(\nu)}=\delta^{-1}t_0^{(\nu)}+\tau_0^{(\nu)}=\rho^{(\nu)}+l^{(\nu)}$
with the complex constants $\rho^{(\nu)}$ and integers $\l^{(\nu)}$, 
$\nu=1,\dots,n,\infty$. Hence $l^{(\nu)}$ and $\delta^{-1}$ are large and 
comparable, $\lim_{\delta\to+0}l^{(\nu)}\,\delta=t_0^{(\nu)}=const$, while 
$\tau_0^{(\nu)}$ remain bounded as $\delta\to+0$. Furthermore, since 
$\delta>0$ and $l^{(\nu)}\in{\Bbb Z}$,
\begin{equation}\label{reality_t0}
t_0^{(\nu)}\in{\Bbb R}. 
\end{equation}
Thus the quantized real part of the vector $\alpha$ of the formal monodromy 
exponents becomes the vector of discrete deformation parameters with the step 
${\cal O}(\delta)$. Redefining $\delta$, we normalize the vector $\Re\alpha$ 
which thus contains $n$ free parameters.

\smallskip
Using (\ref{scaling_change}) in (\ref{Psi_lambda}), (\ref{Psi_x}), we obtain
\begin{eqnarray}\label{Psi_zeta}
&&\frac{d\Psi}{d\zeta}=\delta^{-1}B(\zeta)\Psi,\quad
d\Psi=\omega\Psi,
\\
&&B(\zeta)=\sum_{\nu=1}^n\sum_{k=0}^{r_{\nu}}
\frac{B_{\nu,-k}}{(\zeta-b^{(\nu)}-\delta c^{(\nu)})^{k+1}}-
\sum_{k=1}^{r_{\infty}}B_{\infty,-k}\zeta^{k-1}.\nonumber
\end{eqnarray}
Compatibility of (\ref{Psi_zeta}) reads
\begin{equation}\label{scaled_Frobenius}
dB=[\omega,B]+\delta\frac{\partial\omega}{\partial\zeta},\quad
d\omega=\omega\wedge\omega.
\end{equation}

The method of constructing the approximate solution of the first of
equations (\ref{Psi_zeta}) is comprehensively described in 
\cite{wasow, fedor}. Let ${\Eu R}\subset{\Bbb C}$ be a closed simply 
connected domain satisfying the following conditions: $B(\zeta)$ is 
holomorphic in ${\Eu R}$, and eigenvalues $\mu_j(\zeta)$ of $B(\zeta)$, 
$j=1,2$, have no zero in ${\Eu R}$. This domain can be obtained by removal 
of independent of $\delta$ neighborhoods of all the singularities and turning 
points, i.e.\ zeros of $\mu_1(\zeta)-\mu_2(\zeta)=2\mu(\zeta)$, supplemented 
by certain cuts of the complex $\zeta$-plane. Given $N\in{\Bbb N}$, equation 
(\ref{Psi_zeta}) can be diagonalized in ${\Eu R}$ up to the order 
${\cal O}(\delta^N)$. Integrating the leading diagonal part, one finds the 
{\it WKB approximation\/}, which is holomorphic and invertible in ${\Eu R}$
\cite{fedor},
\begin{eqnarray}\label{WKB}
&&\Psi_{W\!K\!B}(\zeta)=
W(\zeta)\Bigl[I+\sum_{k=1}^{N-1}\delta^kW^{(k)}(\zeta)\Bigr]
\exp\Bigl\{\delta^{-1}\int^{\zeta}\Lambda(\zeta)\,d\zeta\Bigr\},
\\
&&\Lambda(\zeta)=\mu\sigma_3-\delta\,\diag(W^{-1}W_{\zeta})+
\sum_{k=2}^{N-1}\delta^k\Lambda^{(k)}(\zeta),\nonumber
\\\label{B_eigenvalue}
&&W=(W_1,W_2),\quad
BW_j=\mu_jW_j,\quad
j=1,2,\quad
\mu=\mu_1=-\mu_2.
\end{eqnarray}
Here the off-diagonal $W^{(k)}(\zeta)$ and diagonal $\Lambda^{(k)}(\zeta)$ 
matrices are recursively determined by the eigenvector matrix $W(\zeta)$.

The contour $\gamma_j(a,b)\subset{\Eu R}$, $j=1,2$, connecting $a$ and $b$ is 
called the $j$-{\it canonical path\/} if 
$\Re\int_{a}^\zeta\mu_j(\zeta)\,d\zeta$, $\zeta\in\gamma_j(a,b)$, does not 
decrease as $\zeta$ runs from $a$ to $b$ along $\gamma_j(a,b)$. A closed 
simply connected domain ${\cal R}\subset{\Eu R}$ is called the 
$j$-{\it canonical domain} if there exists such a point $a_j$ that 
$\forall b\in{\cal R}$ the contour $\gamma(a_j,b)\subset{\cal R}$ connecting 
$a_j$ and $b$ is homotopy equivalent to a $j$-canonical path. If ${\cal R}$ 
is $j$-canonical for both $j=1$ and $j=2$, it is called the 
{\it canonical domain}. In the canonical domain ${\cal R}$, the WKB 
approximation (\ref{WKB}) approaches the true solution of (\ref{Psi_zeta}) 
with the uniform in $\zeta\in{\cal R}$ accuracy ${\cal O}(\delta^N)$ 
\cite{fedor}.

Any small enough neighborhood of a given point $\zeta_0\in{\Eu R}$ is a 
canonical domain. This domain can be extended as follows. Let 
$\gamma_0(a_1,a_2)\!=\!\{\zeta\!\in\!{\Eu R}\colon
\Im\!\int_{\zeta_0}^{\zeta}\!\mu(\zeta)\,d\zeta\!=\!0\}$ be a segment of the 
{\it anti-Stokes level line} which passes through $\zeta_0$ and has endpoints 
$a_1$ and $a_2$. The numbering of the endpoints is chosen in such a way that 
the oriented contour $\gamma_0(a_1,a_2)$ is the 1-canonical path (i.e.\ for 
$\mu_1=\mu$) while $\gamma_0(a_2,a_1)$ is the 2-canonical path (i.e.\ for 
$\mu_2=-\mu$). Consider an arbitrary point of the segment, 
$\zeta_*\in\gamma_0(a_1,a_2)$, and introduce the contour 
$\gamma_*=\{\zeta\in{\Eu R}\colon
\Re\int_{\zeta^*}^\zeta\mu(\zeta)\,d\zeta=0\}$, which is the segment of the 
{\it Stokes level line}. By construction, the union 
${\cal R}=\cup_{\zeta_*\in\gamma_0(a_1,a_2)}\gamma_*$ of all the contours 
$\gamma_*$ is the canonical domain. 

{\it Remark 3}. Near the non-Fuchsian singularity, the canonical domain
${\cal R}$ can be extended beyond the boundary of ${\Eu R}$ to fill out one
of the Stokes sectors \cite{fedor}.

The approximate solution of the system (\ref{Psi_zeta}) in the canonical 
domain can be constructed multiplying the WKB approximation (\ref{WKB}) in a 
right diagonal matrix 
$Q(\tau)\bigl(I+\sum_{n=1}^{N-1}Q_n(\tau)\delta^n+{\cal O}(\delta^N)\bigr)$ 
independent of $\zeta$. Below, we assume that the diagonal matrix is absorbed 
into the eigenvector matrix $W(\zeta)$, so that the approximate solution of 
the system (\ref{Psi_zeta}) is given by (\ref{WKB}).

\section{Modulation of the spectral curve}

In the canonical domain, integration of the system (\ref{Psi_zeta}) is 
reduced in the leading order w.r.t.\ $\delta$ to the eigenvalue problem. 
Equations (\ref{scaled_Frobenius}) at $\delta=0$ with the eigenvalue problem 
(\ref{B_eigenvalue}) are basic ingredients of the algebro-geometric 
integration of the ``soliton" equations, see \cite{IM, DMN, krichever0, DKN}.
The ``spectral curve" $\Gamma$ is determined by the characteristic equation
\begin{equation}\label{spectral_curve}
\det(B(\lambda)-\mu I)=0.
\end{equation}
In our $2\times2$ traceless case, this is the hyperelliptic curve 
$\mu^2=-\det B(\lambda)$ of genus $g=n+r-2$. The spectral curve of the 
original system (\ref{Psi_lambda}) is clearly not stationary since 
$d(\det A)=-d(A^2)=-A\frac{\partial\Omega}{\partial\lambda}-
\frac{\partial\Omega}{\partial\lambda}A$, but if scaled via 
(\ref{scaling_change}), it varies ``slowly",
\begin{equation}\label{ddetB}
d(\det B)=-\delta\bigl(B\frac{\partial\omega}{\partial\zeta}+
\frac{\partial\omega}{\partial\zeta}B\bigr).
\end{equation}
We adopt the term {\it singular} for the curve $\Gamma$ whose topological 
properties as $\delta\neq0$ differ from that at $\delta=0$, e.g.\ when the 
singularities of (\ref{Psi_zeta}) coalesce, or turning points merge with each 
other or with the singular points. Given a parameterization of the curve
$\Gamma$, we define the {\it discriminant set} in the total space of the 
parameters as the set determining the singular curve.

Let $\Gamma_{as}$ be the limiting $\delta=0$ form of the spectral curve 
$\Gamma$ (\ref{spectral_curve}) for generic equation (\ref{Psi_zeta}),
\begin{equation}\label{mu_as_gen}
\mu_{as}^2=-\lim_{\delta\to0}\det B=\frac{P_{2g+2}(\zeta)}
{\prod_{\nu=1}^n(\zeta-b^{(\nu)})^{2(r_\nu+1)}}, 
\end{equation} 
where $P_{2g+2}(\zeta)$ is a polynomial of degree $2g+2$. Totally, the curve 
(\ref{mu_as_gen}) depends on $2g+n+1$ complex parameters which form the
space ${\cal T}\otimes{\cal D}$. Here, ${\cal T}$ is the space of $g$ complex 
deformation parameters $t$ and of $n$ real normalized deformation parameters 
$\Re\alpha$, see Remark~2. Below, we use the notation $t$ for the combined 
vector $(t,\Re\alpha)$. ${\cal D}$ is the space of $g$ complex invariants and 
of $n+1$ imaginary parts $\Im\alpha$. We call the asymptotic spectral curve 
$\Gamma_{as}$ (\ref{mu_as_gen}) {\it generic} if 
1)~$b^{(\nu)}\neq b^{(\rho)}$ for $\nu\neq\rho$, 
2)~$\det B_{\nu,-r_{\nu}}\neq0$, and, 
3)~all the asymptotic turning points are simple. Two first conditions hold 
true for the generic equation (\ref{Psi_zeta}) with the parameters lying 
apart from the fixed 
singularities of (\ref{Frobenius}) while the third condition can be violated. 
However, because the non-generic curves $\Gamma_{as}$ are localized on a 
union of algebraic surfaces of co-dimension one, a small variation of the 
parameters turns the non-generic spectral curve into generic one. In 
contrast, the generic curve $\Gamma_{as}$ is stable w.r.t.\ the small 
variation of the parameters, and the corresponding curve $\Gamma$ is not 
singular at $\delta=0$.

To illustrate the introduced notions, let us consider the spectral curve 
($g=1$) for the non-generic equation (\ref{Psi_lambda}), (\ref{A_P6}),
\begin{equation}\label{mu6}
\mu^2=\frac{\lambda\alpha^2+H}
{\lambda(\lambda-1)(\lambda-x)},\quad
H=\frac{\bigl(x(x-1)\frac{dy}{dx}-y(y-1)\bigr)^2}{4y(y-1)(y-x)}-\alpha^2y.
\end{equation}
Letting $x=t+\delta\tau$, $\alpha=\delta^{-1}\alpha_0+\alpha_1$, 
$\alpha_0\neq0$, we find that $D=\delta^2H\to D_0$ (\ref{P6_integral}) as 
$\delta\to0$. The asymptotic spectral curve for (\ref{mu6}), 
$\mu_{as}^2(\lambda)=\frac{\alpha_0^2(\lambda-\lambda_0)}
{\lambda(\lambda-1)(\lambda-t)}$, $\lambda_0=-\frac{D_0}{\alpha_0^2}$,
depends on the only deformation parameter $t$. The set of the formal 
monodromy exponents includes the parameter $\alpha_0$, where $\Re\alpha_0=1$. 
The complex module coincides with $\lambda_0$. In the space of pairs 
$(t,\lambda_0)$, the discriminant set for (\ref{mu6}) is given by the union 
of planes $(t=0)\vee(t=1)\vee(\lambda_0=0)\vee(\lambda_0=1)\vee(\lambda_0=t)$.

\smallskip
The generic curve (\ref{mu_as_gen}) satisfies the following condition:

\begin{theorem}\label{Theorem2}
Let the asymptotic spectral curve (\ref{mu_as_gen}) for generic equation
(\ref{Psi_zeta}) be generic at the initial point 
$(t_0,D_0)\in{\cal T}\otimes{\cal D}$. Then there exists an open 
neighborhood\/ ${\cal U}\subset{\cal T}$ of the point $t_0$ such that, for 
any closed path $\ell\in\Gamma_{as}$, 
\begin{equation}\label{Krichever}
J_{\ell}(t,D):=\Re\oint_{\ell}\mu_{as}(\zeta)\,d\zeta=h_{\ell},\quad
\forall t\in{\cal U},\quad
h_{\ell}=J_{\ell}(t_0,D_0).
\end{equation}
\end{theorem}

\proof
Let $\ell\in\Gamma$ be an arbitrary closed contour with the base point 
$\zeta_0$. Consider the solution $\Psi(\zeta)$ of (\ref{Psi_zeta}) as a
multivalued mapping of 
${\Bbb C}\backslash\{b^{(1)}+\delta c^{(1)},\dots,b^{(n)}+\delta c^{(n)}\}$ 
into $SL(2,{\Bbb C})$ and the function $\Psi_{W\!K\!B}(\zeta)$ (\ref{WKB}) as 
a multivalued mapping of the Riemann surface of the curve $\Gamma$ into 
$SL(2,{\Bbb C})$. 

The projection $\pi(\ell)$ of the contour $\ell$ on the complex $\zeta$-plane 
passes through a finite number of the overlapping canonical regions 
${\cal R}_k$, $k=1,\dots,s$, ${\cal R}_{s+1}={\cal R}_1$. In each region 
${\cal R}_k$, the function (\ref{WKB}) gives a uniform in $\zeta$ 
approximation for the exact solutions $\Psi_k(\zeta)$ of the system 
(\ref{Psi_zeta}). The exact solutions for the adjacent canonical regions 
${\cal R}_k$, ${\cal R}_{k+1}$ differ from each other in a right multiplier 
$G_k$ which is independent of $\zeta$ and $t\in{\cal T}$, i.e.\ 
$\Psi_{k+1}(\zeta)=\Psi_k(\zeta)G_k$. Thus 
$$
m_{\ell}(\Psi_{s+1}(\zeta))=\Psi_1(\zeta)M_{\ell}G_1\cdots G_s,$$ 
where $m_{\ell}$ is the operator of the analytic continuation along $\ell$, 
and $M_{\ell}$ is the monodromy matrix for $\Psi_1(\zeta)$ along $\ell$.
Using for $\Psi_k(\zeta)$ the WKB approximations, we find
\begin{equation}\label{oint_of_Lambda}
\exp\Bigl\{\delta^{-1}\oint_{\ell}\Lambda(\zeta)\,d\zeta\Bigr\}=
(I+{\cal O}(\delta))G(\delta).
\end{equation}
Due to continuous dependence of $\Psi_k(\zeta)$ and of the generic curve 
(\ref{mu_as_gen}) on all the parameters, the topology of the curve and of the 
contour $\ell\in\Gamma$ preserve as soon as the point $(t,D)$ is close enough 
to the initial point $(t_0,D_0)$. Therefore the r.h.s.\ of 
(\ref{oint_of_Lambda}) preserves while the deformation parameters $t$ vary in 
a neighborhood of $t_0$. Equating the leading orders of the l.h.s.\ for 
(\ref{oint_of_Lambda}) at the initial point $t_0$ and nearby points $t$, we 
arrive at (\ref{Krichever}).
\endproof

Theorem~\ref{Theorem2} immediately provides us with the following assertions:

\begin{corollary}\label{corollary1}
Let\/ ${\cal U}\subset{\cal T}$ be an open domain and let (\ref{Krichever}) 
holds true. Let the curve $\Gamma_{as}$ be generic at the point 
$t_1\in\partial{\cal U}$. Then there exists an open domain\/ 
${\cal W}\subset{\cal T}$ such that ${\cal U}\subset{\cal W}$, 
$t_1\in{\cal W}$, and (\ref{Krichever}) is valid $\forall t\in{\cal W}$.
\end{corollary}

\begin{corollary}\label{corollary2}
Let\/ ${\cal U},{\cal U}'\subset{\cal T}$ be contiguous open domains and 
$\Sigma\subset\{\partial{\cal U}\cap\partial{\cal U}'\}\neq\emptyset$. Let 
$J_{\ell}(t,D)=h_{\ell}\>$ $\forall t\in{\cal U}$, 
$J_{\ell}(t,D)=h_{\ell}'\>$ $\forall t\in{\cal U}'$, and the integral 
$J_{\ell}(t,D)$ is continuous across $\Sigma$. Then $h_{\ell}=h_{\ell}'$.
\end{corollary}

In accord with the Corollaries~\ref{corollary1} and~\ref{corollary2}, the
equation (\ref{Krichever}) is valid in a subdomain ${\cal U}\subset{\cal T}$ 
bounded by the points where the curve $\Gamma_{as}$ degenerates. In other
words, if ${\cal I}$ is the integral manifold for (\ref{Krichever})
parameterized by $t\in{\cal T}$, then the boundary $\partial{\cal U}$ is the 
projection on ${\cal T}$ of the intersection of ${\cal I}$ with the 
discriminant set of $\Gamma_{as}$. Beyond this boundary, the equation 
(\ref{Krichever}) may be invalid.

Varying the contour $\ell$ in (\ref{Krichever}), we obtain the system of 
equations $J_{\ell_j}=h_j$, $j=1,\dots,2g+n+1$, where the contours $\ell_j$ 
form a homology basis of the Riemann surface of $\Gamma_{as}$. For instance,
taking a small circle $c_{\nu}$ around $\zeta=b^{(\nu)}$ as the integration 
path $\ell$, we obtain the equation for entries of $\Im\alpha$,
\begin{equation}\label{Im_t0=c}
\Im t_0^{(\nu)}=-\frac{h_{c_{\nu}}}{2\pi}=const,\quad
\nu=1,\dots,n,\infty,
\end{equation}
which contains the reality condition (\ref{reality_t0}) as a particular case. 

To discuss (\ref{Krichever}) further, it is convenient to impose the 
conditions (\ref{Im_t0=c}) from the very beginning and to remove $n+1$ small 
circles from the ``sufficient" set of contours. The remaining contours 
$\ell_j$, $j=1,\dots,2g$, form a homology basis of the Riemann surface of the 
algebraic curve $w^2=P_{2g+2}(\zeta)$. 

\smallskip
{\it Remark 4}. We refer to the systems of $2g$ real transcendent equations 
(\ref{Krichever}) over the homology basis of the curve $w^2=P_{2g+2}(\zeta)$ 
as the {\it Krichever's system}. Its particular case with the trivial
r.h.s.\ $h_j=0$, $j=1,\dots,2g$, is called the {\it Boutroux' system}. 

\smallskip
The real dimension $2g$ of ${\cal D}$ coincides with the number of the real 
equations in the Krichever's system (\ref{Krichever}). The functions 
$J_{\ell_j}(t,D)$, $j=1,\dots,2g$, are the independent first integrals of the 
completely integrable Pfaffian system $dJ=0$,
\begin{equation}\label{Pfaffian}
\omega\pmatrix{dD\cr d\bar D\cr}=-\Omega\pmatrix{dt\cr d\bar t\cr},
\end{equation}
where $\omega=\{\omega_{ij},\bar\omega_{ij}\}$ and 
$\Omega=\{\Omega_{ik},\bar\Omega_{ik}\}$ are the matrices of the partial 
derivatives, $\omega_{ij}=\partial J_{\ell_i}/\partial D_j$, 
$\Omega_{ik}=\partial J_{\ell_i}/\partial t_k$, $i,j=1,\dots,g$,
$k=1,\dots,g+n$. Here, the bar means the complex conjugation. The Pfaffian 
system (\ref{Pfaffian}) provides us with the modulation equation in the 
differential form. If contracted to a real plane $\Im t=0$, (\ref{Pfaffian}) 
yields the system which is equivalent to the modulation equations of Whitham 
type. 

The scaling limits in the classical Painlev\'e equations are described
generically by elliptic equations, $g=1$. If $g=1$ and 
$\theta:=\frac{\partial}{\partial D}\mu_{as}(\zeta,t,D)\,d\zeta$ 
is a holomorphic 1-form on the Riemann surface of $\Gamma_{as}$, then the 
classical inequality 
$\Im\Bigl(\overline{\oint_a\theta}\cdot\oint_b\theta\Bigr)>0$ provides us
with the condition $\det\omega\neq0$. Since the curve $\Gamma_{as}$ is 
generic, the integral manifold for the Pfaffian system (\ref{Pfaffian}) 
passing through the initial point $(t_0,D_0)$ is well parameterized by 
$2g+n$ real deformation parameters $t$, $\bar t$, and the Krichever's system 
(\ref{Krichever}) determines a differentiable in the real sense complex 
function $D(t,\bar t)$. Below, we omit the dependence of $D$ on $\bar t$.

In some scaling limit problems, the Painlev\'e function is fixed by the 
choice of the monodromy data rather then by initial conditions. The 
dependence of the unknowns on the additional parameter $\delta$ means that we 
consider a fiber space ${\cal E}=\bigl({\cal P},\pi,B\bigr)$ where $\pi$ is 
the projection of the total space ${\cal P}$ on the base space $B$ of the 
scaling parameter $\delta$. Each fiber ${\cal P}(\delta)$ is the functional 
space of the Painlev\'e functions which is isomorphic to the monodromy 
surface ${\cal M}$ for the associated linear system. The function 
$y(.,\delta)$ appears as a section of the fiber space ${\cal E}$. For 
instance, the sections of the fiber space ${\cal E}$ for the isomonodromy 
scaling limits yield a trivial, ``horizontal" foliation. In this respect, it 
is quite important to relate the quantities $h_{\ell}$ in (\ref{Krichever}) 
with the monodromy data of the associated linear system.

Let us introduce the following notions \cite{fedor}. The {\it Stokes line} is 
a Stokes level line emanating from a turning point. The {\it Stokes graph} is 
the union of all the Stokes lines. The {\it Stokes complex} is a connected 
component of the Stokes graph. The {\it Stokes chain} is a broken line whose 
links are the Stokes lines with the common turning points. We associate 
the singular point to a Stokes complex if there is a Stokes chain which 
approaches or encircles this point. The singular point can be associated to 
more than one Stokes complex. If the Stokes complex approaches or encircles
the only singular point then this singularity is non-Fuchsian and there
exist at least two Stokes lines of the Stokes complex approaching this point.

In our next assertion, we assume the generosity of the monodromy data. The 
assumption is specified in \cite{kapaevP2ell, kapaevP2scl, kap_kit2, nov1, 
kapaevP2P4} for particular asymptotic problems in equations PII and PIV as 
the nontriviality of some of the Stokes multipliers and their combinations. 
Below, the assumption means that the ratios of the fundamental solutions 
normalized in accord with (\ref{Psi_formal}) at the singular points 
associated to the same Stokes complex are neither diagonal nor off-diagonal. 
It is clear that a small admissible variation of the monodromy data, i.e.\ 
the variation preserving the monodromy surface ${\cal M}$, turn the 
non-generic monodromy data into generic ones while generic monodromy data are 
stable w.r.t.\ the small admissible variation.

\begin{theorem}\label{Theorem1}
Let the parameters in generic equation (\ref{Psi_zeta}) lie apart from the 
fixed singularities of (\ref{Frobenius}). Let the corresponding monodromy 
data be generic and do not depend on $\delta$ as $\delta\to+0$. Then, for any 
closed path $\ell\in\Gamma_{as}$, the asymptotic curve satisfies the 
condition 
\begin{equation}\label{Boutroux}
J_{\ell}(t,D):=\Re\oint_{\ell}\mu_{as}(\zeta)\,d\zeta=0.
\end{equation}
\end{theorem}

\proof
We give here a sketch proof. Assume for a moment that for a small circle 
$c_{\nu}$ around $b^{(\nu)}$ taken as the integration path $\ell$ in
(\ref{Krichever}) the equation (\ref{reality_t0}) is violated. Then the 
formula $M_{\nu}S_{\nu}^{(1)}\dots S_{\nu}^{(2r_{\nu})}=
e^{2\pi ix_0^{(\nu)}\sigma_3}$, which relates the formal monodromy exponent 
$x_0^{(\nu)}=\delta^{-1}t_0^{(\nu)}+\tau_0^{(\nu)}$ with the monodromy and 
the Stokes matrices of the fundamental solution normalized at $b^{(\nu)}$ by 
(\ref{Psi_formal}), shows the exponential dependence of the product on 
$\delta^{-1}$ which is a contradiction. 

Let $J_{\ell}=h_{\ell}\neq0$ for a contour $\ell$ which encircles two turning 
points. Thus the Stokes graph has at least two connected components 
${\cal S}_j$  $j=1,2$. Without loss of generality, the domain ${\cal R}$
separating ${\cal S}_j$, $j=1,2$, does not contain neither Stokes lines nor 
singular points, and its boundary consists of two Stokes chains 
$\partial{\cal R}_j\subset{\cal S}_j$, $j=1,2$. Choose the turning points 
$a_1\in\partial{\cal R}_1$ and $a_2\in\partial{\cal R}_2$. By construction, 
$2\Re\int_{a_1}^{a_2}\mu_{as}(\zeta)\,d\zeta=h_{\ell}\neq0$.

First compute the ratio of two fundamental solutions normalized by 
(\ref{Psi_formal}) at the singular points $b^{(j)}$ associated to the Stokes 
complexes ${\cal S}_1$ and ${\cal S}_2$. With this aim, choose a Stokes chain 
from a neighborhood of $b^{(j)}$ to $a_j$, $j=1,2$. Since the subdomain of 
${\cal R}$ bounded by the Stokes level lines is a canonical domain for the 
WKB approximation (\ref{WKB}) \cite{fedor}, we find the ratio as the product 
$$
G_{12}=e^{F_1\sigma_3}E_1
\exp\Bigl\{\delta^{-1}\int_{a_2}^{a_1}\mu_{as}(\zeta)\,d\zeta\sigma_3\Bigr\}
E_2^{-1}e^{-F_2\sigma_3},$$
where $F_{j}=\Bigl\{\delta^{-1}\int_{\zeta_j}^{\zeta}\mu\,d\zeta-
\theta^{(j)}(\lambda)\Bigr\}_{\lambda\to b^{(j)}}$. Modulo the terms which
may behave like a power of $\delta^{-1}$, the matrices $E_j$, $j=1,2$, can be 
represented by the products of the monodromy data of the reduced linear 
equations approximating (\ref{Psi_zeta}) near the turning points and of the
exponents of the phase integrals 
$\exp\bigl\{\delta^{-1}\int_{\zeta_k}^{\zeta_l}\mu\,d\zeta\sigma_3\bigr\}$ 
between the successive turning points of the Stokes chain, see 
\cite{its_nov, kapaevWKB}. Therefore, if neither of the matrices $E_j$ is 
diagonal nor anti-diagonal, then, in the leading order, $G_{12}$ 
exponentially depends on $\delta^{-1}$ which is a contradiction. If both or 
one of $E_j$ is diagonal or anti-diagonal then $\Re F_j\neq0$ for one of 
$j=1,2$. In the latter case, compute the ratio of two fundamental solutions 
normalized at the singular points of the same Stokes complex ${\cal S_j}$, 
$$
G_j=e^{F_j\sigma_3}Ne^{-F_k\sigma_3},$$
where $N$ may behave like a power of $\delta^{-1}$. Due to assumption on 
$G_j$, the matrix $N$ is neither diagonal nor anti-diagonal, thus, in the 
leading order, $G_j$ exponentially depends on $\delta^{-1}$ that is a 
contradiction.
\endproof

\smallskip
{\it Remark 5}. Equation (\ref{Boutroux}) implies that the asymptotic as
$\delta\to0$ Stokes graph for (\ref{Psi_zeta}) is connected.

\smallskip
{\it Remark 6}. The assumption of Theorem~\ref{Theorem1} can be relaxed to 
the condition of boundedness or slow dependence of the monodromy data on 
$\delta^{-1}$.

\smallskip
Absence of the initial condition for the Pfaffian system in the assumptions 
of Theorem~\ref{Theorem1} does not prevent the solvability of 
(\ref{Boutroux}) for $D$. Indeed, the degeneration of the spectral curve due 
to merging of the turning points encircled by the contour $\ell$ allows us to 
find a point $(t_0,D_0)$ on the discriminant set which could serve as the 
initial point for (\ref{Pfaffian}). For equation PII, this idea was used in 
\cite{kapaevP2scl} to prove the existence of the differentiable in the real 
sense function $D(t)$ satisfying (\ref{Boutroux}). There is no principal 
obstacle to extend the proof for other classical Painlev\'e equations. The 
preliminary considerations for PIII and PIV which include the description of 
the corresponding discriminant sets can be found in \cite{kapaevP3scl, 
kapaevP4scl}.

To illustrate the said above, consider again equation (\ref{Psi_zeta}), 
(\ref{A_P6}) and its spectral curve (\ref{mu6}). Assuming (\ref{Boutroux}),
we describe its Stokes graph as the union of three finite Stokes lines 
emanating from $\lambda_0$ and terminating at the simple poles $0,1,t$. The 
discriminant set, which is the union of the planes $t=0$, $t=1$, 
$t=\lambda_0$, $\lambda_0=0$, $\lambda_0=1$, intersects with the integral 
manifold for (\ref{Boutroux}) along the real axis $\Im t=0$, moreover the 
values $t<0$, $0<t<1$ and $t>1$ correspond to the planes $\lambda_0=0$, 
$t=\lambda_0$ and $\lambda_0=1$, respectively. Thus if the initial data for 
(\ref{P6}) at $t=t_0$, $\Im t_0<0$ satisfy the conditions 
$\lambda_0\neq0,1,t_0$, then the asymptotic as $\delta\to0$ Painlev\'e 
function is elliptic in the lower half of the complex $t$-plane and 
degenerates on the real line $\Im t=0$. Whether the asymptotics is elliptic 
or remains degenerate above the real line depends on the chosen initial data, 
or, equivalently, on the corresponding monodromy data.

\section{Scaling limits in PI}

In this section, we present the monodromy deformation approach to the scaling 
limits in the Painlev\'e first equation PI,
\begin{equation}\label{PI}
y_{xx}=6y^2+x.
\end{equation}
Equation PI governs the isomonodromy deformations of 
the linear system (\ref{Psi_lambda}) with
\begin{equation}\label{A_PI}
A(\lambda)=(4\lambda^4+x+2y^2)\sigma_3-
i(4y\lambda^2+x+2y^2)\sigma_2-(2y_x\lambda+\frac{1}{2\lambda})\sigma_1.
\end{equation}
Here $\sigma_3=\bigl({1\ \ 0\atop0\ -1}\bigr)$, 
$\sigma_2=\bigl({0\ -i\atop i\ \ 0}\bigr)$, 
$\sigma_1=\bigl({0\ 1\atop1\ 0}\bigr)$. The set of the monodromy data for 
equation (\ref{A_PI}) consists of the Stokes matrices 
$S_k=\Psi_k^{-1}(\lambda)\Psi_{k+1}(\lambda)$ where 
$\Psi_k(\lambda)\sim\exp\bigl[(\frac{4}{5}\lambda^5+x\lambda)\sigma_3\bigr]$
as $|\lambda|\to\infty$, $\arg\lambda\in\bigl(\frac{\pi}{5}(k-\frac{3}{2}),
\frac{\pi}{5}(k+\frac{1}{2})\bigr)$. The Stokes matrices satisfy the cyclic 
relation $S_1S_2S_3S_4S_5=i\sigma_1$ \cite{kapaevP1}, thus the Painlev\'e 
function set is parameterized by the points of the 2-dimensional complex 
monodromy surface:
\begin{eqnarray}\label{monodromy_surface_PI}
&&\hbox{if}\quad1+s_2s_3\neq0\quad\hbox{then}\quad
s_1=\frac{i-s_3}{1+s_2s_3},\quad
s_4=\frac{i-s_2}{1+s_2s_3},\quad
s_5=i(1+s_2s_3),\nonumber
\\
&&\hbox{if}\quad1+s_2s_3=0\quad\hbox{then}\quad
s_2=s_3=i,\quad 
s_5=0,\quad
s_1+s_4=i.
\end{eqnarray}
Physically interesting solutions of equation (\ref{PI}) have the 
non-oscillating asymptotic behavior $y\sim\pm\sqrt{-x/6}+{\cal O}(x^{-2})$ as 
$x\to-\infty$. Their existence has been proved by Ph.~Holmes and D.~Spence 
\cite{HS}. The solution with the asymptotics 
$y\sim-\sqrt{-x/6}+{\cal O}(x^{-2})$ is unique and corresponds to the values 
$s_2=s_3=0$, $s_1=s_4=s_5=i$, while the solutions with the asymptotics 
$y\sim\sqrt{-x/6}+{\cal O}(x^{-2})$ form a 1-parametric family distinguished 
by the condition $1+s_2s_3=0$ \cite{kapaevP1}. In more details,
\begin{equation}\label{degenerate_as}
y(x)\sim y_-(x)+
a(-x)^{-1/8}e^{-\frac{8}{5}(3/2)^{1/4}(-x)^{5/4}},\quad
x\to-\infty,
\end{equation}
where $y_-(x)=\sqrt{-x/6}+{\cal O}(x^{-2})$ does not contain any free 
parameter, and
\begin{equation}\label{a_PI}
a=-\frac{(2/3)^{1/8}}{2\sqrt{2\pi}}\,\frac{s_1-s_4}{2},\quad
s_1+s_4=i.
\end{equation}
On the complex $x$-plane, the Painlev\'e function for $s_2=s_3=i$, $s_5=0$, 
$s_1+s_4=i$ has the asymptotics (\ref{degenerate_as}) as $|x|\to\infty$, 
$\arg x\in\bigl(\frac{3\pi}{5},\frac{7\pi}{5}\bigr)$ with a piece-wise
constant amplitude $a$ which shows a jump across the negative real axis
\cite{kapaevP1, takei1}; on the rays 
$\arg x=\pm\frac{\pi}{5},\frac{3\pi}{5},\frac{7\pi}{5}$, the asymptotics is 
trigonometric; in the interior of the sectors between the indicated rays, the 
asymptotics is elliptic, see \cite{kapaevP1, kap_kit2}. 

The scaling change of variables
\begin{equation}\label{scl_change1}
x=\delta^{-4/5}(t_0+\delta t),\quad
y=\delta^{-2/5}v,\quad
y_x=\delta^{-3/5}v_t,
\end{equation}
transforms (\ref{PI}) into
\begin{equation}\label{string}
v_{tt}=6v^2+t_0+\delta t.
\end{equation}
The reduced form of (\ref{PI}), i.e.\ equation (\ref{string}) at $\delta=0$, 
is easy to integrate, 
\begin{equation}\label{string_int}
v_t^2=4v^3-g_2v-g_3,\quad
g_2=-2t_0,\quad
g_3=const.
\end{equation}
Thus, at $\delta=0$, $v(t)=\wp(t-\varphi;g_2,g_3)$ is the elliptic function of 
Weierstra\ss. Let $v_0=\sqrt{-t_0/6}$. For the particular value $g_3=-8v_0^3$,
the elliptic function degenerates,
\begin{equation}\label{sh_PI}
v(t)=v_0+\frac{3v_0}{\sinh^2\bigl(\sqrt{3v_0}\,(t-\varphi)\bigr)}.
\end{equation}

Below, we find the asymptotics of the degenerate Painlev\'e function, 
$s_2=s_3=i$, $s_5=0$, $s_1+s_4=i$, when $|s_4|$ is large. Also, we find the 
asymptotics of the ``almost degenerate" Painlev\'e functions when $s_5\neq0$ 
is small. In order to solve these problems, let us introduce the elliptic 
curve
\begin{equation}\label{w2}
w^2=(z-z_1)(z-z_3)(z-z_5)=z^3+\frac{t_0}{2}z+\frac{D_0}{4}.
\end{equation}
The Riemann surface $\Gamma_{as}$ for (\ref{w2}) consists of two complex 
$z$-planes pasted along the cuts $[-\infty,z_5]$ and $[z_1,z_3]$. Let 
$\arg w(z)\to0$ as $z\to+\infty$ on the upper sheet. Let the $a$-cycle 
encircles counter-clockwise the cut $[z_1,z_3]$, and the upper half of the 
$b$-cycle connects $z_5$ and $z_3$. Introduce the integrals over $a$- and 
$b$-cycles
\begin{equation}\label{omega_ab}
\omega_a=\frac{1}{2}\oint_a\frac{dz}{w(z)},\quad
\omega_b=\frac{1}{2}\oint_b\frac{dz}{w(z)},\quad
\tau=\frac{\omega_b}{\omega_a},$$
$$
I_{\ell}=\oint_{\ell}w(z)\,dz,\quad
\Omega_{\ell}=\oint_{\ell}\frac{z\,dz}{w(z)},\quad
\Omega_{\ell}^{(j)}=\Omega_{\ell}-2z_j\omega_{\ell},\quad
\ell\in\{a,b\}.
\end{equation}
Introduce the system of the Krichever's equations for the function
$D_0(t_0)$,
\begin{equation}\label{PI_modulation}
2\,\Re I_a=0,\quad
2\,\Re I_b=h_b,\quad
h_b=-\frac{8}{5}(3/2)^{1/4}.
\end{equation}
Due to triviality of $h_a$, the system (\ref{PI_modulation}) admits the
degeneration of the curve (\ref{w2}), $z_1=z_3=(-t_0/6)^{1/2}$,
$D_0=8(-t_0/6)^{3/2}$. Thus the intersection of the discriminant set with
the integral manifold of (\ref{PI_modulation}) satisfies equation 
$\Re(-t_0)^{5/4}=1$. Among five lines satisfying the latter equation, we 
denote by symbol $\gamma$ the line passing through the point $t_0=-1$ and 
asymptotic to the rays $\arg t_0=\pm\frac{3\pi}{5}$.

Finally, let us introduce the perforated region ${\cal R}(c_0,\epsilon_1)$,
\begin{equation}\label{R_ph}
{\cal R}(c_0,\epsilon_1)=\bigl\{t\in{\Bbb C}\colon\quad
|t|<c_0,\ |t-n\omega_a-m\omega_b|>\epsilon_1>0,\
n,m\in{\Bbb Z}\bigr\}.
\end{equation}

\begin{theorem}\label{Theorem4}
Let $|s_4|\gg1$ and $0<c'\leq|\frac{\textstyle s_1}{\textstyle s_4}|\leq c''$ 
for some constants $c',c''$. Then there exist the positive constant $c_0$ and 
the positive small constants $\epsilon$, $\epsilon_1$ such that, to the right 
of $\gamma$, the Painlev\'e function is given by
\begin{equation}\label{y_right}
y(x)=\delta^{-2/5}\wp\bigl(t-\varphi;-2t_0,-D_0\bigr),\quad
t=\delta^{-1/5}x-\delta^{-1}t_0,
\end{equation}
where $D_0(t_0)$ is determined by (\ref{PI_modulation}) and equations
\begin{eqnarray}\label{delta_t5_via_s14}
&&\delta^{-1}=-\frac{\ln|s_4|}{h_b},
\\\label{t-phi_right}
&&t-\varphi=\delta^{-1}\frac{4}{5}t_0+t+
\frac{\omega_a}{2\pi i}\ln(is_4)-
\frac{\omega_b}{2\pi i}\ln\frac{s_1}{s_4}+
{\cal O}(\delta^{\frac{1}{2}-\epsilon})\in{\cal R}(c_0,\epsilon_1).
\end{eqnarray}
\end{theorem}

\proof
It follows from the integral estimates similar to presented in 
\cite{boutroux, doran-wu_joshi, joshi} that there exists the solution 
$y(x)=\delta^{-2/5}\wp\bigl(t-\varphi;-2t_0,-D_0\bigr)$, $|t|<const$, 
of the Cauchy problem for equation PI with the initial data 
$x_0=\delta^{-4/5}t_0$, $y_0=\delta^{-2/5}v_0$, $y_0'=\delta^{-3/5}v_0'$, 
where $|\delta|\ll1$ and $D_0=v_0'^2-4v_0^3-2t_0v_0$. Using 
(\ref{scl_change1}) and (\ref{scaling_change}) with $\varkappa=-1/5$ (see
Remark~1) in (\ref{Psi_lambda}), (\ref{A_PI}), we obtain equation 
(\ref{Psi_zeta}), $\Psi_{\zeta}\Psi^{-1}=\delta^{-1}B$, where
\begin{equation}\label{B_PI}
B(\zeta)=(4\zeta^4+t_0+2v^2+\delta t)\sigma_3-
i(4v\zeta^2+t_0+2v^2+\delta t)\sigma_2-
(2v_t\zeta+\frac{\delta}{2\zeta})\sigma_1.
\end{equation}
The corresponding spectral curve is given by
\begin{eqnarray}\label{mu_as_PI}
&&\mu^2=\mu_{as}^2+\delta r(\zeta),\quad
\mu_{as}^2=4\zeta^2\bigl(4\zeta^6+2t_0\zeta^2+D_0\bigr)=
16\zeta^2w^2(\zeta^2),
\\\label{D_def}
&&D:=v_t^2-4v^3-2t_0v=D_0+\delta D_1,
\\
&&r(\zeta)=4\zeta^2D_1+8t\zeta^2(\zeta^2-v)+2v_t+\frac{\delta}{4\zeta^2}.
\nonumber
\end{eqnarray}
Theorem~\ref{Theorem2} implies that the system (\ref{Krichever}), 
$2\Re I_a=h_a$, $2\Re I_b=h_b$, holds true, and the leading order $D_0(t_0)$
of the parameter $D$ does not depend on $t$. Thus the definition 
of $D$ (\ref{D_def}) is consistent with the elliptic asymptotic ansatz 
(\ref{string_int}),
\begin{equation}\label{wp_eq}
v_t^2=4v^3+2t_0v+D_0=4w^2(v),\quad
v(t)=\wp(t-\varphi;-2t_0,-D_0).
\end{equation}
To find the phase shift $\varphi$, we apply the isomonodromy deformation 
method \cite{its_nov}. Let us introduce notations $a_3=(B)_{11}$, 
$a_+=(B)_{12}$, $a_-=(B)_{21}$ for the entries of the matrix $B$ (\ref{B_PI}) 
and assume the following:
1)~$|v|,|v_t|\leq c$ for some constant $c>0$, and 2)~zeros of $a_+(\zeta)$, 
or, alternatively, of $a_-(\zeta)$, lie apart from the paths of integration 
below. Let us assume that the parameters $t_0$, $D_0$ determining the
elliptic curve (\ref{w2}) and (\ref{wp_eq}) are such that the asymptotic at
$\delta=0$ Stokes lines for the system (\ref{Psi_zeta}), (\ref{B_PI}),
emanating from $\zeta_1=z_1^{1/2}$ and $\zeta_3=z_3^{1/2}$ form the 
continuous chains
$e^{-i\frac{3\pi}{10}}\infty\to\zeta_1\to e^{-i\frac{\pi}{10}}\infty$,
$e^{-i\frac{\pi}{10}}\infty\to\zeta_1\to\zeta_3\to e^{i\frac{\pi}{10}}\infty$ 
and $e^{i\frac{\pi}{10}}\infty\to\zeta_3\to e^{i\frac{3\pi}{10}}\infty$. 
Then the result of \cite{kapaevWKB} implies the expressions for the Stokes 
multipliers as $\Re(\delta^{-1})\to+\infty$, $|\Im(\delta^{-1})|<const$,
\begin{eqnarray}\label{s14}
&&\hskip-15pt
\ln s_4=2\delta^{-1}
\Bigl[\int_{\zeta_1^{(\sigma)}}^{\zeta}\nu_{\sigma}\,d\zeta-
\theta+
\sigma\frac{\delta}{2}\ln\frac{a_{\sigma}}{\mu}
\Bigr]_{\zeta\to e^{-i\frac{\pi}{10}}\infty}-\sigma\ln(-2i)+
{\cal O}(\delta),\nonumber
\\
&&\hskip-15pt
\ln s_1=2\delta^{-1}
\Bigl[\int_{\zeta_3^{(\sigma)}}^{\zeta}\nu_{\sigma}\,d\zeta-
\theta+
\sigma\frac{\delta}{2}\ln\frac{a_{\sigma}}{\mu}
\Bigr]_{\zeta\to e^{i\frac{3\pi}{10}}\infty}-\sigma\ln(-2i)+
{\cal O}(\delta),
\end{eqnarray}
where $\zeta_k^{(\sigma)}$ are zeros of $\nu_{\sigma}(\zeta)$,
$\zeta_k^{(\sigma)}\bigr|_{\delta=0}=z_k^{1/2}$, $k=1,3$,
\begin{equation}\label{nu0}
\nu_{\sigma}^2=\mu_{as}^2+\delta r+
\sigma\delta\bigl(a_3'-a_3\frac{a_{\sigma}'}{a_{\sigma}}\bigr),\quad
\theta=\frac{4}{5}\zeta^5+(t_0+\delta t)\zeta,\quad
\sigma\in\{+,-\}.
\end{equation}
Since the Stokes multipliers do not depend on $t$, and, at the critical point 
$t=t_5$, $v_t=0$, $v=z_5$, the assumptions above are not violated for the 
non-degenerate curve (\ref{w2}), we may compute the r.h.s.\ of (\ref{s14})
at this point. Using the change $\zeta^2=z$, we transform the elliptic 
integrals in (\ref{s14}) as follows (look for similar details in 
\cite{kapaevP2ell, kapaevP2scl}):
\begin{eqnarray}\label{s14_expand_at_5}
&&\ln(-is_4)=-2\delta^{-1}I_b-\frac{t_5}{2}\Omega_b^{(5)}-
\frac{D_1(t_5)}{2}\omega_b+
{\cal O}(\delta^{\frac{1}{2}-\epsilon}),\nonumber
\\
&&\ln\frac{s_1}{s_4}=-2\delta^{-1}I_a-\frac{t_5}{2}\Omega_a^{(5)}-
\frac{D_1(t_5)}{2}\omega_a+
{\cal O}(\delta^{\frac{1}{2}-\epsilon}),
\end{eqnarray}
where $\epsilon>0$ is small. The conditions $\ln|s_4|\gg1$ and
$c'\leq|\frac{\textstyle s_1}{\textstyle s_4}|\leq c''$ are consistent with 
(\ref{s14_expand_at_5}) if $h_a=0$ and $h_b<0$. Substitutions 
$\delta\mapsto\alpha^{5/4}\delta$, $t_0\mapsto\alpha t_0$, 
$t\mapsto\alpha^{-1/4}t$, $v\mapsto\alpha^{1/2}v$, 
$v_t\mapsto\alpha^{3/4}v_t$, which leave $x$, $y$ and $y_x$ invariant, allow 
us to demand for $h_b$ the value in (\ref{PI_modulation}). The conditions on 
the Stokes graph imposed above are consistent with the reality condition,
$D_0(t_0)\in{\Bbb R}$ for $t_0>-1$. The latter implies that the chosen branch 
of $D_0(t_0)$ yields degeneration of the curve (\ref{w2}) on $\gamma$, and 
the validity domain of the elliptic asymptotic solution is located to the 
right of $\gamma$.

Using in (\ref{s14_expand_at_5}) the identities 
\begin{equation}\label{identities}
\omega_aI_b-\omega_bI_a=\frac{4\pi i}{5}t_0,\quad
\omega_a\Omega_b-\omega_b\Omega_a=4\pi i,
\end{equation}
which follow from the Legendre identity, we find the critical point $t_5$,
$$
t_5=-\delta^{-1}\frac{4}{5}t_0-
\frac{\omega_a}{2\pi i}\ln(-is_4)+
\frac{\omega_b}{2\pi i}\ln\frac{s_1}{s_4}+
{\cal O}(\delta^{\frac{1}{2}-\epsilon}),$$
and the expression for the argument of the $\wp$-function of Weierstrass
(\ref{t-phi_right}).
\endproof

\smallskip
{\it Remark 7}. Because $\omega_a$ and $\omega_b$ are periods of the elliptic 
function, the asymptotics (\ref{y_right}) is invariant w.r.t.\ the choice of 
the argument of the Stokes multipliers.

\begin{theorem}\label{Theorem5}
Let $0<|s_5|\ll1$, $0<c'\leq|s_2|\leq c''$ for some constants $c',c''$. Then 
there exist positive constant $c_0$ and positive small constants $\epsilon$, 
$\epsilon_1$ such that, to the left of $\gamma$, the Painlev\'e function is 
given by (\ref{y_right}) where $D_0(t_0)$ is determined by 
(\ref{PI_modulation}) and
\begin{eqnarray}\label{delta_t3_via_s25}
&&\delta^{-1}=\frac{\ln|s_5|}{h_b},
\\\label{t-phi_left}
&&t-\varphi=\delta^{-1}\frac{4}{5}t_0+t-
\frac{\omega_a}{2\pi i}\ln(is_5)+
\frac{\omega_b}{2\pi i}\ln(is_2)+
{\cal O}(\delta^{\frac{1}{2}-\epsilon})\in{\cal R}(c_0,\epsilon_1).
\end{eqnarray}
\end{theorem}

\proof
In essential, we repeat the proof of Theorem~\ref{Theorem4}. Existence of the 
elliptic asymptotic solution of the Cauchy problem follows from the integral 
estimates similar to presented in \cite{boutroux, doran-wu_joshi, joshi}. 
Using (\ref{scl_change1}) in (\ref{Psi_lambda}), (\ref{A_PI}), we obtain 
equation (\ref{Psi_zeta}), $\Psi_{\zeta}\Psi^{-1}=\delta^{-1}B$, with the 
matrix $B$ (\ref{B_PI}) whose spectral curve is given by (\ref{mu_as_PI}). 
Applying Theorem~\ref{Theorem2}, we arrive at the system (\ref{Krichever}) 
for the function $D_0(t_0)$, $2\Re I_a=h_a$, $2\Re I_b=h_b$. The definition 
of $D_0$ (\ref{D_def}) is consistent with the elliptic asymptotic ansatz 
(\ref{string_int}), (\ref{wp_eq}). Assume that $|v|,|v_t|\leq c$ and zeros of 
$a_+(\zeta)$ or $a_-(\zeta)$ lie apart from the paths of integration below. 
Let us assume also that the parameters $t_0$, $D_0$ determining the elliptic 
curve (\ref{w2}) and (\ref{wp_eq}) are such that the asymptotic at $\delta=0$ 
Stokes lines for the system (\ref{Psi_zeta}), (\ref{B_PI}), emanating from 
$\zeta_k=z_k^{1/2}$, $k=1,3,5$, form the continuous chains 
$e^{-i\frac{\pi}{10}}\infty\to\zeta_1\to e^{i\frac{\pi}{10}}\infty$,
$e^{i\frac{\pi}{10}}\infty\to\zeta_1\to\zeta_3\to e^{i\frac{3\pi}{10}}\infty$ 
and $e^{i\frac{3\pi}{10}}\infty\to\zeta_5\to e^{i\frac{\pi}{2}}\infty$. 
Then, using the result of \cite{kapaevWKB}, we find the Stokes multipliers as 
$\Re(\delta^{-1})\to+\infty$, $|\Im(\delta^{-1})|<const$,
\begin{eqnarray}\label{s25}
&&\hskip-15pt
\ln s_2=-2\delta^{-1}
\Bigl[\int_{\zeta_5^{(\sigma)}}^{\zeta}\nu_{\sigma}\,d\zeta-
\theta+
\sigma\frac{\delta}{2}\ln\frac{a_{\sigma}}{\mu}
\Bigr]_{\zeta\to e^{i\frac{\pi}{2}}\infty}+\sigma\ln(2i)+{\cal O}(\delta),
\\
&&\hskip-15pt
\ln s_5=-2\delta^{-1}
\Bigl[\int_{\zeta_1^{(\sigma)}}^{\zeta}\nu_{\sigma}\,d\zeta-
\theta+
\sigma\frac{\delta}{2}\ln\frac{a_{\sigma}}{\mu}
\Bigr]_{\zeta\to+\infty}+\sigma\ln(2i)+{\cal O}(\delta),\quad
\sigma\in\{+,-\}.\nonumber
\end{eqnarray}
We compute the r.h.s.\ of (\ref{s25}) at the critical point $t=t_3$, $v_t=0$, 
$v=z_3$ since the Stokes multipliers do not depend on $t$ and the assumptions 
above are not violated for the non-degenerate curve (\ref{w2}). Using the
method of \cite{kapaevP2ell, kapaevP2scl}, we transform the elliptic integrals above 
as follows,
\begin{eqnarray}\label{s25_expand_at_3}
&&\ln(-is_2)=2\delta^{-1}I_a+\frac{t_3}{2}\Omega_a^{(3)}+
\frac{D_1(t_3)}{2}\omega_a+
{\cal O}(\delta^{\frac{1}{2}-\epsilon}),\nonumber
\\
&&\ln(-is_5)=2\delta^{-1}I_b+\frac{t_3}{2}\Omega_b^{(3)}+
\frac{D_1(t_3)}{2}\omega_b+
{\cal O}(\delta^{\frac{1}{2}-\epsilon}),
\end{eqnarray}
where $\epsilon>0$ is small. The condition $|s_5|\ll1$ and the boundedness of 
$s_2$ are consistent with (\ref{s25_expand_at_3}) if $h_a=0$ and $h_b<0$.
Substitutions $\delta\mapsto\alpha^{5/4}\delta$, $t_0\mapsto\alpha t_0$,
$t\mapsto\alpha^{-1/4}t$, $v\mapsto\alpha^{1/2}v$,
$v_t\mapsto\alpha^{3/4}v_t$, which leave $x$, $y$ and $y_x$ invariant, allow
us to scale $h_b$ to the value in (\ref{PI_modulation}). The conditions on 
the Stokes graph imposed above are consistent with $D_0(t_0)\in{\Bbb R}$ for 
$t_0<-1$. Thus, it is chosen the branch of $D_0(t_0)$ which yields the
degeneration of the curve (\ref{w2}) on $\gamma$, and the validity domain of 
the elliptic asymptotic solution is located to the left of $\gamma$.

Using in (\ref{s25_expand_at_3}) the identities (\ref{identities}), we find 
the critical point $t_3$,
$$
t_3=-\delta^{-1}\frac{4}{5}t_0
+\frac{\omega_a}{2\pi i}\ln(-is_5)-
\frac{\omega_b}{2\pi i}\ln(-is_2)+
{\cal O}(\delta^{\frac{1}{2}-\epsilon}),$$
and the argument of the $\wp$-function of Weierstrass (\ref{t-phi_left}).
\endproof

\smallskip
The description of the Painlev\'e function on the line $\gamma$ is given by
the following assertion.

\begin{theorem}\label{Theorem6}
Let $t_0\in\gamma$ where the asymptotic spectral curve (\ref{mu_as_PI}) 
degenerates via $z_1=z_3$. Let the functions $v,v_t$ introduced in 
(\ref{scl_change1}) satisfy the estimates 
$v-z_1,v_t={\cal O}(\delta^{\varepsilon})$, 
$\varepsilon\in\bigl[0,\frac{1}{2}\bigr]$. Then the Stokes multipliers of the 
associated linear system (\ref{Psi_lambda}), (\ref{A_PI}) are as follows:
\begin{description}
\item{i)} if $0\leq\varepsilon<\frac{1}{3}$ and 
$v_t-\hat\sigma2(v-z_1)\sqrt{v+2z_1}={\cal O}(\delta^{1-\varepsilon})$,
$\hat\sigma^2=1$,
\begin{eqnarray}\label{s401b}
&&s_4=-i\frac{\sqrt{v+2z_1}-\hat\sigma\sqrt{3z_1}}
{\sqrt{v+2z_1}+\hat\sigma\sqrt{3z_1}}
\sqrt{2\pi}\frac{e^{i\frac{\pi}{2}\rho_b}}{\Gamma(\frac{1}{2}+\rho_b)}
e^{F_b}\bigl(1+
{\cal O}(\delta^{\frac{1}{4}-\frac{3}{4}\varepsilon})\bigr),\nonumber
\\
&&s_5=-i\frac{\sqrt{v+2z_1}+\hat\sigma\sqrt{3z_1}}
{\sqrt{v+2z_1}-\hat\sigma\sqrt{3z_1}}
\sqrt{2\pi}\frac{e^{i\frac{\pi}{2}\rho_b}}{\Gamma(\frac{1}{2}-\rho_b)}
e^{-F_b}\bigl(1+
{\cal O}(\delta^{\frac{1}{4}-\frac{3}{4}\varepsilon})\bigr),
\\
&&s_1=e^{-2\pi i\rho_b}s_4
\bigl(1+{\cal O}(\delta^{1-\frac{\varepsilon}{2}})\bigr),\nonumber
\\
&&1+s_4s_5=-e^{2\pi i\rho_b}
\bigl(1+{\cal O}(\delta^{1-\frac{\varepsilon}{2}})\bigr),\quad
\arg(-1-s_4s_5)\in\bigl(-\frac{2\pi}{3},\frac{2\pi}{3}\bigr),\nonumber
\end{eqnarray}
where
\begin{equation}\label{F_b}
F_b=\delta^{-1}\frac{8}{5}(3/2)^{\frac{1}{4}}(-t_0)^{\frac{5}{4}}+
\rho_b\ln\bigl[
i\delta^{-1}2^{\frac{19}{4}}3^{\frac{5}{4}}(-t_0)^{\frac{5}{4}}\bigr]-
2(3/2)^{\frac{1}{4}}(-t_0)^{\frac{1}{4}}t,
\end{equation}
$$
\hskip-4pt
\rho_b=\frac{1}{4\sqrt{3z_1}}
(\delta^{-1}(v_t^2-4(v-z_1)^2(v+2z_1))-2t(v-z_1)+\frac{v_t}{v-z_1}),\
\Re\rho_b\in\bigl(-\frac{1}{3},\frac{1}{3}\bigr);$$
\item{ii)} if $\frac{1}{3}<\varepsilon\leq\frac{1}{2}$ then
\begin{eqnarray}\label{s401c}
&&\hskip-16pt
s_4=\frac{3^{1/8}2^{7/8}(-t_0)^{1/8}\delta^{1/2}}{v_t-2\sqrt{3z_1}(v-z_1)}
\sqrt{2\pi}\frac{e^{i\frac{\pi}{2}\rho_c}e^{-i\pi/4}}
{\Gamma(\rho_c)}e^{F_c}\bigl(1+
{\cal O}(\delta^{\frac{1}{4}-\frac{3}{8}\varepsilon})+
{\cal O}(\delta^{3\varepsilon-1})\bigr),\nonumber
\\
&&\hskip-16pt
s_5=\frac{3^{1/8}2^{7/8}(-t_0)^{1/8}\delta^{1/2}}{v_t+2\sqrt{3z_1}(v-z_1)}
\sqrt{2\pi}\frac{e^{i\frac{\pi}{2}\rho_c}e^{-i\pi/4}}
{\Gamma(-\rho_c)}e^{-F_c}\bigl(1+
{\cal O}(\delta^{\frac{1}{4}-\frac{3}{8}\varepsilon})+
{\cal O}(\delta^{3\varepsilon-1})\bigr),\nonumber
\\
&&\hskip-16pt
s_1=-e^{-2\pi i\rho_c}s_4
\bigl(1+{\cal O}(\delta^{1-\frac{\varepsilon}{2}})\bigr),
\\
&&1+s_4s_5=e^{2\pi i\rho_c}
\bigl(1+{\cal O}(\delta^{1-\frac{\varepsilon}{2}})\bigr),\quad
\arg(1+s_4s_5)\in\bigl(-\frac{\pi}{3},\frac{\pi}{3}\bigr),\nonumber
\end{eqnarray}
where
\begin{equation}\label{F_c}
F_c=\delta^{-1}\frac{8}{5}(3/2)^{1/4}(-t_0)^{5/4}+
\rho_c\ln(i\delta^{-1}2^{19/4}3^{5/4}(-t_0)^{5/4})-
2\sqrt{3z_1}\,t+i\frac{\pi}{4},
\end{equation}
$$
\rho_c=\frac{\delta^{-1}}{4\sqrt{3z_1}}\bigl(v_t^2-12z_1(v-z_1)^2\bigr)+
{\cal O}(\delta^{3\varepsilon}),\quad
\Re\rho_c\in\bigl(-\frac{1}{6},\frac{1}{6}\bigr).$$
\end{description}
\end{theorem}

In other notations, this assertion was obtained in \cite{kapaevP1} following 
the conventional arguments of the isomonodromy deformation method 
\cite{its_nov}. Applying the method of \cite{kapaevWKB}, it is possible to
justify the expressions above. It is also possible to improve the estimates 
for the error terms. To avoid the repetition and to save space, we skip this 
known derivation.

The inversion of (\ref{s401b}) yields the asymptotics of the Painlev\'e 
function (\ref{sh_PI}) with $v_0=z_1=\sqrt{-t_0/6}$ and the phase given in 
terms of the Stokes multipliers by
\begin{equation}\label{sh_PI_via_s45}
\sqrt{3v_0}\,(t-\varphi)=-\frac{F_b}{2}-\frac{1}{2}\ln A,\quad
A=-i\sqrt{2\pi}
\frac{e^{i\frac{\pi}{2}\rho_b}}{s_4\Gamma(\frac{1}{2}+\rho_b)},
\end{equation}
where $F_b$ is defined in (\ref{F_b}), and
$$
\rho_b=\frac{1}{2\pi i}\ln(-1-s_4s_5),\quad
\arg(-1-s_4s_5)\in\bigl(-\frac{2\pi}{3},\frac{2\pi}{3}\bigr).$$

For our purposes, the inversion of (\ref{s401c}) is more interesting. In the
leading order, this yields
\begin{eqnarray}\label{v_lin}
&&v=\sqrt{-t_0/6}+\delta^{1/2}\bigl(Ae^{-F_c}-Be^{F_c}\bigr),
\\
&&AB=3^{-1/4}2^{-7/4}(-t_0)^{-1/4}\rho_c
\bigl(1+
{\cal O}(\delta^{\frac{1}{4}-\frac{3}{8}\varepsilon})+
{\cal O}(\delta^{3\varepsilon-1})\bigr),\nonumber
\\
&&\rho_c=\frac{1}{2\pi i}\ln(1+s_4s_5)
+{\cal O}(\delta^{1-\frac{\varepsilon}{2}}),\quad
\arg(1+s_4s_5)\in\bigl(-\frac{\pi}{3},\frac{\pi}{3}\bigr),\nonumber
\\
&&A=3^{-1/8}2^{-7/8}(-t_0)^{-1/8}\sqrt{2\pi}
\frac{e^{i\frac{\pi}{2}\rho_c}e^{-i\pi/4}}{s_5\Gamma(-\rho_c)}\bigl(1+
{\cal O}(\delta^{\frac{1}{4}-\frac{3}{8}\varepsilon})+
{\cal O}(\delta^{3\varepsilon-1})\bigr),\nonumber
\\
&&B=3^{-1/8}2^{-7/8}(-t_0)^{-1/8}\sqrt{2\pi}
\frac{e^{i\frac{\pi}{2}\rho_c}e^{-i\pi/4}}
{s_4\Gamma(\rho_c)}\bigl(1+
{\cal O}(\delta^{\frac{1}{4}-\frac{3}{8}\varepsilon})+
{\cal O}(\delta^{3\varepsilon-1})\bigr),\nonumber
\end{eqnarray}
and $F_c$ is defined in (\ref{F_c}).

Now, we are prepared to discuss the problems announced earlier.

\smallskip
{\it i) The large amplitude separatrix solution.}

Let us consider the Painlev\'e function $y(x)$ corresponding to the Stokes 
multipliers (\ref{monodromy_surface_PI}) satisfying $1+s_2s_3=0$, i.e.\ 
$s_2=s_3=i$, $s_5=0$, $s_1+s_4=i$, when $|s_4|\gg1$. The asymptotics 
(\ref{degenerate_as}), (\ref{a_PI}) explain the used term the 
{\it large amplitude separatrix solution}. Introduce $t_0$, $t$ and $v$ as in 
(\ref{scl_change1}) and consider the boundary $\gamma$ for the domain of
validity of (\ref{PI_modulation}). Then applying Theorem~\ref{Theorem4}, we 
see that to the right of $\gamma$, the asymptotics of the Painlev\'e function 
is elliptic and is described by the equations (\ref{y_right}), 
(\ref{delta_t5_via_s14}), (\ref{t-phi_right}). The corresponding elliptic  
curve is determined by the system of the Krichever's equations 
(\ref{PI_modulation}).

On $\gamma$, the elliptic curve degenerates, and, since $1+s_4s_5=1$, the
asymptotics of the Painlev\'e function is described by the limiting as
$\rho_c\to0$ form of (\ref{v_lin}),
\begin{eqnarray}\label{v_deg_lin}
&&v=\sqrt{-t_0/6}+\delta^{1/2}Ae^{F},\quad
\delta^{-1}=-\frac{\ln|s_4|}{h_b},\quad
\quad
h_b=-\frac{8}{5}(3/2)^{1/4},\nonumber
\\
&&A=3^{-1/8}2^{-7/8}(-t_0)^{-1/8}
\frac{s_4}{\sqrt{2\pi}}\bigl(1+{\cal O}(\delta^{1/16})\bigr),
\\
&&F=-\delta^{-1}\frac{8}{5}(3/2)^{1/4}(-t_0)^{5/4}+
2^{3/4}3^{1/4}(-t_0)^{1/4}t.\nonumber
\end{eqnarray}
To the left of $\gamma$, the curve (\ref{w2}) remains degenerate, and the
Krichever's system is not applicable. Observing however, that the complete 
series for the solution with the leading terms (\ref{v_deg_lin}) contains the 
positive degrees of $e^F$ only, we may conjecture that the asymptotics of the 
Painlev\'e function to the left of $\gamma$, i.e.\ for $\Re(-t_0)^{5/4}>1$, 
is given by the analytic continuation of (\ref{v_deg_lin}). This conjecture 
can be confirmed by the use of the Riemann-Hilbert problem approach, see 
\cite{DZ, its_fok_kap}. An alternative investigation of such series based on 
the Borel summation method is given in \cite{takei1, costin}.

\smallskip
{\it ii) The small perturbation of the degenerate solution.}

Let $0<|s_5|\ll1$, 
$c'\leq|s_2|,\bigl|\frac{\textstyle s_1}{\textstyle s_4}\bigr|\leq c''$ for
some positive finite constants $c',c''$, e.g., when $s_2=i(1+\epsilon)$, 
$s_3=i(1+\epsilon\varkappa)$, for $0<|\epsilon|\ll1$, 
$c'\leq|\varkappa|\leq c''$. Using Theorem~\ref{Theorem5}, we see that, to 
the left of the line $\gamma_l=\gamma$ defined above, the asymptotics of 
$y(x)$ is described by (\ref{y_right}), (\ref{delta_t3_via_s25}), 
(\ref{t-phi_left}). The elliptic curve (\ref{w2}) is determined by the 
Krichever's system (\ref{PI_modulation}) with $h_a=0$, 
$h_b=-\frac{8}{5}(3/2)^{1/4}$. 

Due to Theorem~\ref{Theorem4}, the right boundary $\gamma_r$ of the 
discriminant set consists of the points satisfying the equation 
$\Re(-t_0)^{5/4}=-\frac{\ln|{\textstyle s_4}|}{\ln|{\textstyle s_5}|}<1$ and 
is asymptotic to the rays $\arg t_0=\pm\frac{3\pi}{5}$. If $s_4$ is bounded as
$\epsilon\to0$, the right boundary $\gamma_r$ coincides with these rays. To 
the right of $\gamma_r$, the asymptotic solution $y(x)$ is given by 
(\ref{y_right}), (\ref{t-phi_right}) where the elliptic curve (\ref{w2}) is 
determined by the Krichever's system with the parameters $h_a'=0$ and 
$h_b'=-h_b\frac{\ln|{\textstyle s_4}|}{\ln|{\textstyle s_5}|}$. If $s_4$ is 
bounded, then $h_a'=h_b'=0$, and the description of the Painlev\'e function 
can be obtained by the use of changes (\ref{scl_change1}) in the formulae of 
\cite{kapaevP1, kap_kit2} for $y(x)$ as $|x|\to\infty$. On the boundaries 
$\gamma_l$ and $\gamma_r$, we may use (\ref{v_lin}) with 
$\delta^{-1}=\frac{\ln|{\textstyle s_5}|}{h_b}$, 
$h_b=-\frac{8}{5}(3/2)^{1/4}$, $\rho_c\to0$. As easy to see, both the 
non-constant terms in (\ref{v_lin}) decrease as $\epsilon\to0$ for 
$-\frac{\ln|{\textstyle s_4}|}{\ln|{\textstyle s_5}|}<\Re(-t_0)^{5/4}<1$. 
Because the complete expansion of the Painlev\'e function involves the 
positive degrees of these terms only, we may conjecture the validity of 
(\ref{v_lin}) between $\gamma_l$ and $\gamma_r$ as well.

In particular, the discussion above provides us with the possibility to give 
the analytic interpretation for the formal 2-parametric ``instanton type" 
series with the initial terms given by (\ref{v_lin}) and extensively studied 
in \cite{kawai_takei, takei1, takei2}.

\medskip
{\bf Acknowledgments.} This work was partially supported by RFBR under grant
number 99--01--00687. The author is grateful to a referee for his valuable
comments and to A.~Kitaev for remarks.

\ifx\undefined\bysame
\newcommand{\bysame}{\leavevmode\hbox to3em{\hrulefill}\,}
\fi

\end{document}